\documentclass[12pt,dvipsnames]{article}
 
\usepackage{iftex}
\ifLuaTeX
  \usepackage{polyglossia} 
  \usepackage{fontspec} 
  \setmainlanguage{english}
  \usepackage[nosort]{cite}
\else 
  \usepackage[utf8]{inputenc}
  \usepackage[T1]{fontenc}
  \usepackage[nosort]{cite}
  \usepackage[english]{babel} 
\fi

\usepackage[top=2cm,bottom=2cm,outer=2.5cm,inner=2.5cm,marginparwidth=2.5cm]{geometry}
\usepackage{fvextra} 
\usepackage{authblk} 

\ifLuaTeX
  \usepackage[unicode]{hyperref}
  \usepackage{pdftexcmds} 
  \makeatletter
  \let\pdfstrcmp\pdf@strcmp
  \makeatother
\else
  \usepackage{hyperref}
\fi

\usepackage{amsmath}
\usepackage{amsfonts}
\usepackage{amssymb}

\usepackage{lmodern} 
\usepackage{titlesec} 
\usepackage{fancyhdr} 
\usepackage{abstract} 

\usepackage{xcolor}

\hypersetup{
pdfstartview={FitH},
pdftitle={M2-doughnuts},
pdfauthor={N. Drukker, M. Trépanier},
pdfsubject={},
pdfcreator={},
pdfkeywords={},
colorlinks=true,
citecolor=MidnightBlue,
linkcolor=MidnightBlue,
urlcolor=MidnightBlue
}

\numberwithin{equation}{section}

\newcommand\frontmatter{%
  \clearpage
  \pagenumbering{roman}
}

\newcommand\mainmatter{%
  \clearpage
  \pagenumbering{arabic}
}

\newcommand{\vev}[1]{\left\langle #1 \right\rangle}
\newcommand{\diff}{\mathrm{d}}

\def\bH {\mathbb{H}}

\def\bR {\mathbb{R}}

\def\bZ {\mathbb{Z}}

\def\aM{{\mathsf{M}}}

\def\aR{{\mathsf{R}}}
\def\aQ{{\mathsf{Q}}}
\def\aS{{\mathsf{S}}}

\def\cN{{\mathcal{N}}}

\def\cL{{\mathcal{L}}}
\def\cO{{\mathcal{O}}}

\newcommand{\bea}{\begin{eqnarray}}
\newcommand{\eea}{\end{eqnarray}}
\newcommand{\beq}{\begin{equation}}
\newcommand{\eeq}{\end{equation}}
\newcommand{\bal}{\begin{equation}\begin{aligned}}
\newcommand{\eal}{\end{aligned} \end{equation}}

\title{M2-doughnuts}
\author[1]{Nadav Drukker\thanks{\href{mailto:nadav.drukker@gmail.com}{nadav.drukker@gmail.com}}}
\author[2]{Maxime
  Tr\'epanier\thanks{\href{mailto:trepanier.maxime@gmail.com}{trepanier.maxime@gmail.com}}}
\affil[1]{\it Department of Mathematics, King's College London,\protect\\London, WC2R
2LS, United Kingdom}
\affil[2]{\it Institute of Theoretical and Mathematical Physics,\protect\\
Moscow State University (ITMP), Moscow 119991, Russia}
\date{}

\begin{document}

\frontmatter
\maketitle
\thispagestyle{empty}

\begin{abstract}
We present a family of new M2-brane solutions in $AdS_7\times S^4$ that calculate toroidal BPS 
surface operators in the $\cN=(2,0)$ theory. These observables 
are conformally invariant and not subject to anomalies so we are 
able to evaluate their finite expectation values at leading order at large $N$.
In the limit of a thin torus we find a cylinder, which is a natural surface generalization 
of both the circular and parallel lines Wilson loop. We study and comment on this limit in some detail.
\end{abstract}

\mainmatter

\section{Introduction and conclusions}
\label{sec:intro}

The $AdS$/CFT correspondence \cite{maldacena:1997re} allows to calculate
nonperturbative field theory quantities using classical geometry. A case in
point are surface operators of the 6d $\cN=(2,0)$ theory, which are described at
large $N$ in terms of classical M2-branes in $AdS_7 \times S^4$
\cite{maldacena:1998im}. There is a clear analogy with the calculation of Wilson
loops in $\cN=4$ supersymmetric Yang-Mills theory in 4d, but while there are
plenty of examples of classical string solutions in $AdS_5\times S^5$ (see for
example 
\cite{Drukker:1999zq, berenstein:1998ij, Gubser:2002tv, Mikhailov:2002ya, 
Drukker:2005cu, Hofman:2006xt, Alday:2007hr, Kruczenski:2014bla, Huang:2016atz, He:2017cwd
}), explicit M2-brane solutions are very rare: The main examples are a pair of
anti-parallel planes \cite{maldacena:1998im} and the sphere
\cite{berenstein:1998ij} (see also~\cite{chen:2008ds} for a generalisation of
the anti-parallel planes to finite temperature).

For arbitrary surfaces, the asymptotic form of the solution near the boundary of
$AdS$ was calculated and this is sufficient to evaluate their anomaly
\cite{graham:1999pm, Drukker:2020dcz}.  Recall that two dimensional surface
observables generally suffer from conformal anomalies, which are generalizations
of the usual central charge in 2d CFT and dimensions of local operators
\cite{deser:1993yx,Boulanger:2007st,Schwimmer:2008yh,solodukhin:2008dh,Drukker:2020dcz}.

As usual, these anomalies arise from divergences in the evaluation of physical
quantities and are local, which is why the asymptotic solution is enough to
capture them. They also normally render the exact normalisation of a physical
operator ambiguous. This is analogous to the prescription dependence in defining
the normalisation of local operators as the number $C$ in
$\vev{\cO(x)\cO(y)}\sim C/|x-y|^{2\Delta}$.  In recent years a lot of progress
has been achieved in the exact evaluation of the coefficients entering the
anomaly of surface operators of the $\cN = (2,0)$ theory 
\cite{Gentle:2015jma,Rodgers:2018mvq,Jensen:2018rxu,Estes:2018tnu,Chalabi:2020iie,Wang:2020xkc}, 
so the time is now ripe to turn to evaluating finite expectation values for
``anomaless'' surfaces, the cases where the anomaly vanishes.

The recent paper \cite{Drukker:2020bes} presented a whole slew of surface
operators in the $\cN = (2,0)$ theory that are BPS.
In all of the classes of operators studied, invariance under
supersymmetry implies that the corresponding M2-brane is calibrated with respect
to appropriate calibration forms. This translates into first-order differential
equations for their embedding in $AdS$ and drastically simplifies the task of
finding new classical solutions.  Furthermore, there are quite a few examples of
anomaless surfaces, so their expectation values should be well defined.

The purpose of this note is to present new nontrivial classical M2-brane
solutions which also have a nontrivial expectation value.  The surfaces we
evaluate here are flat tori in $\bR^4$ (so can also be thought of as Clifford
tori in $S^3$) and break the $R$-symmetry $SO(5)\to SO(3)$ (they couple to 2
scalar fields in the abelian version of the theory). In the language of
\cite{Drukker:2020bes}, they belong to type-$\bH$, or more specifically, to
sub-type-L of Lagrangian surfaces in $\bR^4$ and they preserve 4
supercharges.

In the following we present the classical solutions for any such torus with
radii $R_1$ and $R_2$. The final expression for the expectation value of the
surface operator is \eqref{eqn:action}
\beq
\label{eqn:Sren}
\vev{V_{R_1,R_2}}=e^{-S_\text{ren}}\,,
\qquad
S_\text{ren}
=-\pi N\left(\frac{R_1}{R_2}+\frac{R_2}{R_1}\right).
\eeq
As we argue below, this combination of the cross-ratio $R_1/R_2$ and a constant are the 
only allowed functions compatible with the symmetries of the torus. So what we really 
determined is the prefactor $-\pi$ and absence of constant piece. 
A simple calculation in the abelian theory, relying on the techniques of 
\cite{henningson:1999xi,gus03,gustavsson:2004gj,Drukker:2020dcz} gives a trivial answer. 
Note that there is no Lagrangian description of the 6d theory for finite $N$, and no other
calculation of this quantity. 
One approach to finite $N$ would be to perform a semiclassical M-theory analysis beyond 
the leading large $N$ limit considered here.

An even simpler version of this observable is the limit of $R_1\to\infty$ with fixed $R_2$, 
studied in Section~\ref{sec:cylinder}. The torus becomes a cylinder, where the invariant cross-ratio 
is now the ratio of its length $2\pi R_1$ to its radius $R_2$, as in \eqref{Scylren}.

This observable shares some similarities to the quark-antiquark potential in a gauge theory. The latter 
is represented by the two antiparallel lines Wilson loop observable. One natural uplift to surface observables 
is to replace each of the lines with a plane, as was done in \cite{maldacena:1998im}. Another one is to 
replace the $\bR\times \bZ_2$ geometry with $\bR\times S^1$, hence giving the cylinder.

What we study here is a version of this observable that preserves a fraction of the supercharges, so 
is BPS, yet like the circular Wilson loop in $\cN=4$ SYM, it has a nontrivial vacuum expectation value 
\cite{erickson:2000af, drukker:2000rr, pestun:2007rz}. 
The factor $-N/2R_2$ in \eqref{Scylren} is then one natural analog to the quark-antiquark potential in this 
theory at large $N$. The value we find here is very different from that arising in the two plane calculation in 
\cite{maldacena:1998im}, as they are very distinct analogues to the quark-antiquark observable.

We hope to report on more finite results for BPS and non-BPS surface observables in the $\cN=(2,0)$ theory 
in the near future.

\section{The tori}
\label{sec:tori}

In the absence of a Lagrangian description, surface operators in the $\cN =
(2,0)$ theory are defined indirectly 
by stating their properties: their geometry, the local breaking of R-symmetry (also referred to as 
``scalar coupling'', in analogy with Maldacena-Wilson loops, though there are no fields) 
and a representation of the $ADE$ algebra underlying the theory. 
This is discussed in great detail in~\cite{Drukker:2020bes} which also analyses 
possible relationships between the geometry and scalar couplings that guarantee preserved 
supersymmetry. That analysis is similar to that of~\cite{ zarembo:2002an, drukker:2007qr ,Dymarsky:2009si} 
in the case of Wilson loops in $\cN=4$ SYM.

One of the classes of BPS surface operators identified in~\cite{Drukker:2020bes}, named 
type-$\bH$, consists of operators of arbitrary geometry in $\bR^4\subset\bR^6$ and the supersymmetry condition is 
that the scalar coupling (or local $SO(5)$ vector) is
\beq
n^I = \frac{1}{2} \eta^{I}_{\mu\nu} \varepsilon^{ab} \partial_a x^\mu
\partial_b x^\nu\,,
\label{eqn:typeHansatz}
\eeq
where $\eta^{I}_{\mu\nu}$ is the chiral 't~Hooft symbol. A convenient choice is
\begin{align}
  \eta_{14}^1 = \eta_{23}^1 = \eta_{24}^2 = -\eta_{13}^2 = \eta_{12}^3 =
  \eta_{34}^3 = 1\,,
\end{align}
with the other components fixed by symmetry or zero. 
Note that $\varepsilon^{ab}$ in \eqref{eqn:typeHansatz} is normalized with a factor
of the induced metric on the surface such that $\|n\|=1$.

The surface operators we study here are particular examples of operators of type-$\bH$. They 
are flat tori in $\bR^4$ and we parametrise their embedding by $\varphi_1, \varphi_2 \in[0,2\pi)$ and
fixed $R_1, R_2$ as
\bal
\label{torus}
x^1 &= R_1 \cos{\varphi_1}\,, \qquad&
x^2 &= R_1 \sin{\varphi_1}\,,\\
x^3 &= R_2 \cos{\varphi_2}\,, \qquad&
x^4 &= R_2 \sin{\varphi_2}\,.
\eal
Plugging this into \eqref{eqn:typeHansatz} implies the scalar couplings should be
\beq
n^1 = -\sin(\varphi_1+\varphi_2)\,, \qquad
n^2 = \cos(\varphi_1+\varphi_2)\,.
\label{nI}
\eeq
Supersymmetry does not depend on the choice of representation, but 
we focus on operators in the fundamental representation of $A_{N-1}$, which
by the AdS/CFT dictionnary are dual to M2-branes at large
$N$~\cite{maldacena:1998im}.

The analysis of~\cite{Drukker:2020bes} shows that generic surfaces of type-$\bH$ preserve 
a single Poincar\'e supercharge $\aQ$ (out of 16 $\aQ$'s and 16 $\aS$'s). 
It is also shown there that for Lagrangian surfaces 
in $\bR^4$, where $n^3=0$, these supercharges are doubled. Furthermore, because 
the torus is in $S^3$, the tori also preserve two 2 special supercharges $\aS$'s. 

A concrete way to understand the supersymmetries is to notice that our ansatz
has two $U(1)$ symmetries, respectively the shifts in $\varphi_1$ and
$\varphi_2$. The preserved supercharges are those invariant under the action of
these two $U(1)$ symmetries (with $\aM_{\mu\nu}$ are rotations and $\aR_{IJ}$ are
R-symmetry transformations)
\begin{align}
  \left[ \aM_{12} - \aR_{12}, \aQ \right] = 
  \left[ \aM_{34} - \aR_{12}, \aQ \right] = 0\,.
  \label{eqn:preservedQ}
\end{align}
Using the 6d (anti)chiral gamma matrices $\gamma_\mu$ ($\bar{\gamma}_\mu$) and
the 5d R-symmetry gamma matrices $\check{\gamma}_I$, the supercharges satisfying
these constraints can be constructed explicitly, with
\begin{align}
  \label{eqn:preservedQcond}
  (1 + \gamma_1 \bar{\gamma}_2 \gamma_3 \bar{\gamma}_4) \aQ
  = (1 - i \gamma_1 \bar{\gamma}_3 \check{\gamma}_1) \aQ
  = (1 - i \gamma_1 \bar{\gamma}_4 \check{\gamma}_2) \aQ
  = 0\,.
\end{align}
These are 3 independent constraints, so two $\aQ$ out of 16 satisfy them.
The first says that $\aQ$ is antichiral in 4d, which can also be seen by taking the
difference of the equations~\eqref{eqn:preservedQ}.
For more details and the analysis of special supersymmetries $\aS$,
see~\cite{Drukker:2020bes}.

Note that because the supercharges are antichiral, their anticommutator only sees a
combination of the two $U(1)$ symmetries
\beq
\left\{ \aQ, \aS \right\} \propto (\aM_{12} + \aM_{34} - 2\aR_{12})\,,
\eeq
with all other commutators vanishing. This is compatible with the discrete
symmetry of the tori where we exchange $\varphi_1$ with $\varphi_2$ and $R_1$ with $R_2$,
and in the following we assume without loss of generality that $R_1 \ge R_2$.

Beside their supersymmetry, another important property of these operators is
that they do not suffer from conformal anomalies, so their expectation value is
well-defined. The reason is that the anomalies 
of generic operators of type-$\bH$ are related to the Euler class and self-intersection 
number of the surface (see details in~\cite{Drukker:2020bes}), but in the case of 
the torus those two topological invariants vanish.

Note that there are other BPS tori that belong to the type-S class in
\cite{Drukker:2020bes}. They have different $n^I$ and are not studied here.

\section{Doughnut solutions}
\label{sec:doughnut}

We now find the classical solution of M2-branes in $AdS_7\times S^4$ ending on the tori presented above. 
It is convenient to use the metric
\beq
ds^2=4L^2 y\,\diff x^\mu\diff x^\mu
+\left(\frac{L}{y}\right)^2\diff y^I\diff y^I\,,
\label{eqn:metric}
\eeq
with $\mu=1,\cdots,6$, $I=1,\cdots,5$ and $y=\|y^I\|$. Note also that compared to more common forms of 
this metric we absorbed a factor of $L^3$ into $y^I$. 
The background also has 
$F_4$ and $F_7$ fluxes, but those do not play a role in the calculation below.

We look for our M2-brane to have $U(1)^2$ isometry related to $\aM_{12}  - \aR_{12}$ and 
$\aM_{34} - \aR_{12}$. 
Choosing as worldvolume coordinates 
the angles $\varphi_1,\varphi_2$ and $\rho = \sqrt{y_1^2 + y_2^2}$, the ansatz
is%
\footnote{To avoid confusion with exponents we use subscripts $y_I$ instead of
  superscripts to denote the components of $y$.}
\begin{equation}
\label{ansatz}
  \begin{gathered}
    x^1 = r_1(\rho) \cos\varphi_1\,, \qquad
    x^2 = r_1(\rho) \sin\varphi_1\,,\\
    x^3 = r_2(\rho) \cos\varphi_2\,,\qquad 
    x^4 = r_2(\rho) \sin\varphi_2\,,\\
    y_1 = -\rho \sin(\varphi_1+\varphi_2)\,, \qquad
    y_2 = \rho \cos(\varphi_1+\varphi_2)\,, \qquad
    y_3(\rho)\,.
  \end{gathered}
\end{equation}
This ansatz is clearly in an $AdS_5\times S^2$ subspace, but in fact because the boundary surface is a Clifford 
torus in $S^3$, there is a further restriction to $AdS_4$ and the solutions should satisfy the constraint
\beq
\label{ads4}
r_1^2 + r_2^2 + \frac{1}{\sqrt{\rho^2 + y_3^2}}= R_1^2 + R_2^2\,.
\eeq

The boundary of space is at $y\to\infty$, and we expect $y_3/\rho\to0$ there (because $n^3=0$ in 
\eqref{nI}). Indeed we see that $y_1$ and $y_2$ in \eqref{ansatz} match the expectation from \eqref{nI}. 
Likewise we should impose $r_1\to R_1$ and $r_2\to R_2$. At the interior of space the 
surface should end at $\rho=0$, such that $y_1=y_2=0$ and with either $r_1=0$ or $r_2=0$.

The Lagrangian for the M2-brane is
\beq
\cL=2L^3
\frac{\sqrt{\big[4 \left((r_1')^2+(r_2')^2 \right)(\rho^2+y_3^2)^{3/2}+1+(y_3')^2\big]
\big[4r_1^2r_2^2(\rho^2+y_3^2)^{3/2}+ \left( r_1^2 + r_2^2 \right) \rho^2\big]}}{(\rho^2+y_3^2)^{3/4}}
\,.
\eeq
It is natural to assume that for the BPS solution the two terms in the square root 
are proportional to each other. This means either
\beq
\frac{4(r_1')^2+4(r_2')^2}{1+(y_3')^2}=
\frac{4r_1^2r_2^2}{(r_1^2+r_2^2)\rho^2}\,,
\qquad\text{or}\qquad
\frac{4(r_1')^2+4(r_2')^2}{1+(y_3')^2}=
\frac{r_1^2+r_2^2}{4r_1^2r_2^2}\frac{\rho^2}{(\rho^2+y_3^2)^3}\,.
\eeq
The second one looks more natural and can be rewritten as
\beq
4(r_1')^2+4(r_2')^2=
\left(\frac{1}{r_1^2}+\frac{1}{r_2^2}\right)\frac{\rho^2(1+(y_3')^2)}{4(\rho^2+y_3^2)^3}\,.
\eeq
This is easily solved with
\beq
\label{BPSeqn1}
2r_i'=\frac{1}{2r_i}\frac{\rho\sqrt{1+(y_3')^2}}{(\rho^2+y_3^2)^{3/2}}\,.
\eeq

We can get a third equation by differentiating the constraint \eqref{ads4}. Substituting the 
equations above to eliminate $r_i'$, we find an equation for $y_3$ of the form
\beq
(\rho^2-y_3^2)(y_3')^2=-2\rho y_3 y_3'\,.
\eeq
The only solutions to this equation with $y_3/\rho\to0$ at large $\rho$ and no poles at finite 
$\rho$ are with $y_3'=0$.

Plugging this into \eqref{BPSeqn1}, we can immediately integrate to get
\bal
\label{r1r2}
r_1^2(\rho)&=R_1^2-\frac{1}{2\sqrt{\rho^2+y_3^2}}
=R_1^2-\frac{1}{2y}\,,
\\
r_2^2(\rho)&=R_2^2-\frac{1}{2\sqrt{\rho^2+y_3^2}}
=R_2^2-\frac{1}{2y}\,.
\eal
The integration constants $R_1^2$ and $R_2^2$ are fixed from the boundary conditions 
at $y\to\infty$, where we recover $r_1\to R_1$ and $r_2\to R_2$. 
The only remaining unknown is the constant $y_3$, and that can 
be determined from the minimum value of $y$ at $\rho=0$, where either $r_1(0)=0$ or $r_2(0)=0$. 
As we assume that $R_2 \le R_1$, the first circle to shrink is $r_2$, so this happens 
at $\rho=0$ if $y_3=\pm\frac{1}{2R_2^2}$. The different signs give two different solutions, 
where the M2-brane wraps the lower or upper hemisphere of the $S^2\subset S^4$. In the following 
we use the positive choice, in order not to write $|y_3|$ everywhere.

These expressions indeed solve the equations of motion and are derived more rigorously in the next section from the calibration 
equations.

The classical value of the Lagrangian is now
\beq
\cL=L^3
\left(4r_1r_2+\frac{r_1^2+r_2^2}{r_1r_2}\frac{\rho^2}{(\rho^2+y_3^2)^{3/2}}\right)\,.
\eeq
It is convenient to change variables to $y$, where
\beq
\cL'=\cL\,\frac{\diff\rho}{\diff y}=
2 L^3\frac{(2y_3 (R_1^2 + R_2^2)-1) |y|^3+(y_3 (R_1^2+R_2^2)-1) (y+y_3)y_3y-y_3^3}
{y_3 |y|^2 \sqrt{(y+y_3) ((2y_3 (R_1^2+R_2^2)-1) y-y_3)}}\,.
\eeq
Integrating over $y$, $\varphi_1$ and $\varphi_2$ we find
\beq
\int \cL'\,\diff y\,\diff \varphi_1\,\diff \varphi_2
=8\pi^2 L^3\frac{(y-y_3)\sqrt{(y_3+y)((2 y_3 (R_1^2+R_2^2)-1) y-y_3)}}{y_3 y}
\eeq
This vanishes at $y=y_3$ and diverges for $y\to\infty$. Expanding in that limit and 
multiplying by the M2-brane tension
\beq
T_{M2}=\frac{1}{4\pi^2\ell_P^3}=\frac{N}{4\pi L^3}\,,
\eeq
gives
\begin{equation}
\label{eqn:action}
  \begin{aligned}
    S
    &=8\pi^2 L^3T_{M2}y\frac{\sqrt{(2 y_3 (R_1^2+R_2^2)-1)}}{y_3}
    -8\pi^2 L^3T_{M2}\frac{y_3(R_1^2 +R_2^2)}{\sqrt{(2 y_3 (R_1^2+R_2^2)-1)}}\\
    &=4\pi N y R_1 R_2
    -\pi N \left( \frac{R_1}{R_2} + \frac{R_2}{R_1} \right).
  \end{aligned}
\end{equation}
The divergent term is $\frac{yN}{\pi}\text{Area}(\text{torus})$ and is cancelled 
by an appropriate counterterm.
We are left with a finite conformally invariant answer \eqref{eqn:Sren}.

\section{BPS equations}
\label{sec:BPS}

The derivation above was rather intuitive, but involved guesswork. To put it on 
firmer footing we resort to the BPS equations presented in \cite{Drukker:2020bes}. In the case of surfaces of 
type-$\bH$, the M2-branes are calibrated with respect to the 3-form
\beq
\label{eqn:typeH3form}
\phi=
2L^3 \eta_{\mu\nu}^I \diff x^{\mu}\wedge\diff x^\nu\wedge \diff y^I
-\frac{L^3}{y^3}\diff y_1\wedge\diff y_2\wedge\diff y_3\,,
\eeq
and obey the differential equations
\beq
  \partial_\alpha X^M =
  \frac{1}{2} g_{\alpha\beta} \varepsilon^{\beta\gamma\delta} G^{ML} \phi_{LNP}
  \partial_\gamma X^N \partial_\delta X^P\,.
  \label{eqn:g2structurePDE}
\eeq
Here $X^M$ are any of $x^\mu$ and $y^I$, 
$g$ is the induced metric, $\varepsilon$ is
the Levi-Civita tensor (with $1/\sqrt{g}$) and 
$G$ is the target space metric.

These equations are simple to solve. Let us examine the components of
$\phi_{LNy_3}$ and restrict to our $AdS_4\times S^2$ with coordinates as in
\eqref{ansatz}: $r_1$, $r_2$, $\varphi_1$, $\varphi_2$, $\rho$, $y_3$.  We have
\bal
\label{y3phi}
\partial_{y_3}\cdot\phi
&= 4L^3 \left( \diff x^1\wedge \diff x^2+\diff x^3\wedge \diff x^4 \right) -\frac{L^3}{y^3}\diff y_1\wedge\diff y_2
\\
&= 4L^3 \left( r_1\diff r_1\wedge \diff \varphi_1 +r_2\diff r_2\wedge \diff
\varphi_2
-\frac{\rho}{4(\rho^2+y_3^2)^{3/2}}\diff \rho\wedge(\diff\varphi_1+\diff\varphi_2) \right)
\,.
\eal
Recall that our ansatz \eqref{ansatz} 
assumes $\partial_{\varphi_1} y_3=\partial_{\varphi_2} y_3=0$ (this also follows
from the supersymmetry preserved).
Taking $X^M$ in 
\eqref{eqn:g2structurePDE} to be $y_3$ we find
\bal
\partial_{\varphi_1} y_3
&\propto r_2r_2'-\frac{\rho}{4(\rho^2+y_3^2)^{3/2}}\,,
\\
\partial_{\varphi_2} y_3
&\propto 
r_1r_1'-\frac{\rho}{4(\rho^2+y_3^2)^{3/2}}\,.
\eal
Setting those to zero exactly reproduces the equations \eqref{BPSeqn1} with $y_3'=0$. Indeed, 
we can also derive $y_3'=0$ from the fact that there is no $\diff \varphi_1
\wedge\diff \varphi_2$ terms on the second 
line on \eqref{y3phi}.

We thus rederived the equations in Section~\ref{sec:doughnut} from the BPS conditions 
and consequently their solution \eqref{r1r2}.

\section{Cylinder limit}
\label{sec:cylinder}

In the discussion above the torus is defined for any two radii $R_1$ and
$R_2$. In the limit where $R_1 \to \infty$ and $R_2$ is kept fixed, the torus
locally becomes a cylinder of radius $R_2$. More precisely, redefining $v=R_1\varphi_1$ and
shifting $x^1\to x^1-R_1$, the torus \eqref{torus} becomes
\bal
x^1 &= 0\,, \qquad&
x^2 &= v\,,\\
x^3 &= R_2 \cos{\varphi_2}\,, \qquad&
x^4 &= R_2 \sin{\varphi_2}\,.
\eal
This is a cylinder of radius $R_2$ extended along the $x^2$ direction. The scalar couplings \eqref{nI}
are
\beq
n^1 = -\sin\varphi_2\,, \qquad
n^2 = \cos\varphi_2\,.
\eeq

A cylinder is clearly embedded in an $\bR^3$ subspace of $\bR^4$ so according to the classification 
of \cite{Drukker:2020bes}, is of type-N. Being extended uniformly in the $x^2$ direction, it is also 
a version of the type-$\bR$ surfaces, which are the uplift to 6d of the BPS Wilson loops of 
\cite{zarembo:2002an}. Taking the limit does not change the number of preserved 
supercharges, but replaces the two special supersymmetries $\aS$ with two further $\aQ$. 
To see that, note that in 3d there is no
chirality condition (first condition of~\eqref{eqn:preservedQcond}) so the
number of $\aQ$ is doubled.

Taking the same limit on the M2-brane ansatz \eqref{ansatz} and solution \eqref{r1r2}, 
we find the limiting solution
\bal
x^1 &= 0\,, \quad&
x^2 &= v\,,\quad&
x^3&=r_2(\rho)\cos\varphi_2\,,\quad&
x^4&=r_2(\rho)\sin\varphi_2\,,\\
&&y_3&=\frac{1}{2R_2^2}\,,\quad&
y_1&=-\rho\sin\varphi_2\,,\quad&
y_2&=\rho\cos\varphi_2\,,
\eal
with
\beq
r_2(y)=\sqrt{R_2^2-\frac{1}{2y}}\,,
\qquad
y=\sqrt{\rho^2+\frac{1}{4R_2^4}}\,.
\eeq

The action can also be derived as a limit of the action in \eqref{eqn:action}. We find
\beq
S
=4\pi N yR_1 R_2
-\frac{\pi NR_1}{R_2}\,.
\eeq
Identifying the first term with $\frac{Ny}{\pi} \text{Area}(\text{cylinder})$, we see 
that the length $D$ of the cylinder is represented by $D=2\pi R_1$. Thus we interpret the last term as
\beq
\label{Scylren}
S_\text{ren}
=-\frac{ND}{2R_2}\,.
\eeq
This has the natural structure arising from a conformal potential, extensive in the length $D$ and inversely 
proportional to the size $2R_2$.

This interpretation also places an interesting restriction on the expectation
value of the torus~\eqref{eqn:Sren}. Using conformal symmetry and the discrete
symmetry exchanging $R_1 \leftrightarrow R_2$, the expectation value can be an
arbitrary function of 
$\frac{R_1}{R_2} + \frac{R_2}{R_1}$. But
because it has to be extensive with $R_1$ in the limit $R_1 \to \infty$, the
only possibilities are
\beq
  S_\mathrm{torus} = s_0 + s_1 \left( \frac{R_1}{R_2} + \frac{R_2}{R_1}
  \right),
\eeq
with the same constant $s_1$ appearing in the expectation value of the cylinder.

As mentioned above, the cylinder reduces under compactification to $\cN=4$ SYM to the BPS Wilson loops of 
\cite{zarembo:2002an}, which famously have vanishing expectation values 
\cite{Guralnik:2003di,Guralnik:2004yc, Dymarsky:2006ve}. Within the classification of \cite{Drukker:2020bes}, 
such surfaces in the $\cN=(2,0)$ theory are of type-$\bR$. It was verified there that in the abelian 
theory they indeed have vanishing expectation values. Contrary to the expectation in that paper, we see here 
that this does not carry over to the holographic theory, as we found a nonzero
answer.

In $\cN=4$ SYM the loops of \cite{zarembo:2002an} have trivial expectation value and the interesting observable 
is the 1/2 BPS circular Wilson loop with fixed scalar coupling \cite{erickson:2000af, drukker:2000rr, pestun:2007rz}. 
The cylinder with fixed scalar coupling is not BPS in the $\cN=(2,0)$ theory. Instead, the BPS observables 
are the sphere with 
fixed coupling \cite{berenstein:1998ij}, or the cylinder studied here. The sphere is subject to 
conformal anomalies, but the cylinder does not, giving a new finite BPS observable in the theory.

Indeed, all the BPS tori discussed in this paper and their large $N$ M2-doughnut avatars are finite BPS 
observables. It would be interesting to understand whether there are corrections subleading in $N$ 
augmenting \eqref{eqn:action} and \eqref{Scylren}.

\section*{Acknowledgements}
We are grateful Arkady Tseytlin 
 for helpful communications.
ND's research is supported by the Science Technology \& Facilities council under the grants
ST/T000759/1.

\bibliographystyle{utphys2}
\bibliography{ref}

\end{document}